# Phonon Quasi-Particles and Anharmonic Free Energy in Complex Systems

Dong-Bo Zhang,[1] Tao Sun,[1] and Renata M. Wentzcovitch[1,2]

[1]*Department of Chemical Engineering and Materials Science and*

[2]*Minnesota Supercomputing Institute,*

*University of Minnesota, Minneapolis, MN 55455*

## Abstract

We use a hybrid strategy to obtain anharmonic frequency shifts and lifetimes of phonon quasi-particles from first principles molecular dynamics simulations in modest size supercells. This approach is effective irrespective of crystal structure complexity and facilitates calculation of full anharmonic phonon dispersions, as long as phonon quasi-particles are well defined. We validate this approach to obtain anharmonic effects with calculations in $MgSiO_3$-perovskite, the major Earth forming mineral phase. First, we reproduce irregular temperature induced frequency shifts of well characterized Raman modes. Second, we combine the phonon gas model (PGM) with quasi-particle frequencies and reproduce free energies obtained using thermodynamic integration. Combining thoroughly sampled quasi-particle dispersions with the PGM we then obtain first-principles anharmonic free energy in the thermodynamic limit ($N \rightarrow \infty$).

Thermodynamic and thermoelastic properties of minerals are of crucial importance in geophysics providing the basis for large scale geodynamic simulations [1] and mineralogical interpretations of seismological data [2]. Because experiments at pressures and temperatures of the Earth's interior are challenging and often involve large uncertainties, theoretical calculations of these properties are particularly valuable [3]. A paradigm for computing thermal properties of crystalline materials is the phonon gas model (PGM) [4-7] which uses the phonon spectrum to obtain vibrational entropy and other thermodynamic quantities. The PGM works well in general, even for strongly anharmonic systems insofar as phonon quasi-particles exist, i.e., have well defined frequencies and lifetimes [5]. For nearly harmonic systems, a simplification can be made by assuming phonon frequencies have no intrinsic temperature dependence and depend on volume only. This so-called quasi-harmonic approximation (QHA) [8,9] combined with harmonic phonon spectra determined by first principles [10] has been successfully applied to a large variety of Earth minerals, including $MgSiO_3$-perovskite (MgPv) [11-13] the most abundant phase of the Earth's mantle. Despite this success, intrinsic anharmonic effects have been identified in the Raman spectrum of MgPv at ambient pressure [14-16], in the olivine to wadsleyite transition boundary [3] - the cause of the major 410 km seismic discontinuity - and in the high temperature stabilization of other mantle phases, e.g., cubic $CaSiO_3$-perovskite (CaPv) [17] and Mg-ortho-enstatite [18]. Therefore, there is a clear need to advance calculations of anharmonic phonon spectra and free energy for mineral physics applications. First principles anharmonic phonon calculations so far have focused on relatively simple crystal structures [19-23]. The complexity of Earth forming silicates motivates further advances in this area.

In this Letter we use a hybrid scheme that combines first principles molecular dynamics (MD) and harmonic phonon calculations to characterize phonon quasi-particles and compute anharmonic free energies. We obtain intrinsic temperature frequency shifts and lifetimes of all modes sampled in the MD simulation by analyzing power spectra of mode-projected velocity-autocorrelation functions (VAF). This scheme is applicable not only to weakly anharmonic systems such as MgPv [13], but also to strongly anharmonic systems, such as cubic CaPv [17,24], where some harmonic phonons have imaginary frequencies at ambient conditions. Crystal structure complexity poses no limitations to this scheme.

The usual VAF for crystalline systems is defined as [25],

$$\left\langle V_{\mathbf{q}}(0)V_{\mathbf{q}}(t)\right\rangle = \lim_{\tau\to\infty}\frac{1}{\tau}\int_0^\tau V_{\mathbf{q}}(t')V_{\mathbf{q}}(t'+t)\,\mathrm{d}\,t', \qquad (1)$$

where $\mathbf{q}$ are Brillouin zone wave-vectors of the primitive cell sampled in the MD simulation, $V_{\mathbf{q}}(t)=\sum_{i=1}^{N}V(t)\cdot e^{i\mathbf{q}\cdot\mathbf{R}_i}$ is the $\mathbf{q}$-projected velocity, and $\mathbf{R}_i$ is the atomic coordinate of atom $i$ in the supercell. $V(t)=\left\{v_1(t)\sqrt{M_1},...,v_N(t)\sqrt{M_N}\right\}$ with $3N$ components is the weighted velocity and $v_i(t)$ (i=1,…,$N$) are atomic velocities produced by MD simulations with $N$ atoms per supercell. Its power spectrum,

$$G_{\mathbf{q}}(\omega)=\int_0^\infty \left\langle V_{\mathbf{q}}(0)V_{\mathbf{q}}(t)\right\rangle e^{i\omega t}\,\mathrm{d}\,t, \qquad (2)$$

characterizes the collective dynamics of modes with wave-vector $\mathbf{q}$, including anharmonic effects. Simple crystals have only few non-equivalent modes with widely different frequencies well separated in $G_{\mathbf{q}}(\omega)$ [25]. Analysis of $G_{\mathbf{q}}(\omega)$ can offer well defined frequencies and line-widths of modes with the same $\mathbf{q}$. This approach has been successfully used to compute phonon lifetimes and the thermal conductivity of MgO [20], a simple mineral with the rocksalt structure. However, complex crystals have numerous phonon modes with similar frequencies. Their individual spectra overlap and cannot be resolved in $G_{\mathbf{q}}(\omega)$.

This issue can be resolved by calculating separately the power spectra of individual modes with wave-vector $\mathbf{q}$ and polarization branch $\mathbf{s}$ ($3N$ in total),

$$G_{\mathbf{q,s}}(\omega)=\int_0^\infty \left\langle V_{\mathbf{q,s}}(0)V_{\mathbf{q,s}}(t)\right\rangle e^{i\omega t}\,\mathrm{d}\,t, \qquad (3)$$

where

$$V_{\mathbf{q,s}}(t)=\sum_{i=1}^{N}V(t)\exp(-i\mathbf{q}\cdot\mathbf{R}_i)\cdot\hat{\mathbf{e}}_{\mathbf{q,s}}, \qquad (4)$$

with $\hat{\mathbf{e}}_{\mathbf{q,s}}$ being the polarization vector of normal mode $(\mathbf{q},\mathbf{s})$ of the static crystal in the equilibrium state. This strategy was used in conjunction with modified embedded atom model potentials to investigate the vibrational density of state (VDOS) of silicon nano-particles and correlate it with the bulk VDOS [26]. The $(\mathbf{q},\mathbf{s})$-projected VAF, $\langle V_{\mathbf{q},\mathbf{s}}(0)V_{\mathbf{q},\mathbf{s}}(t)\rangle$, can also be used to investigate anharmonic effects since it describes the dynamics of mode $(\mathbf{q},\mathbf{s})$ interacting with other modes sampled in the MD simulation. For a well-defined phonon quasi-particle, $\langle V_{\mathbf{q},\mathbf{s}}(0)V_{\mathbf{q},\mathbf{s}}(t)\rangle$ displays oscillatory decaying behavior that can be described by a renormalized temperature dependent phonon frequency, $\tilde{\omega}_{\mathbf{q},\mathbf{s}}$ , and a line-width, $\Gamma_{\mathbf{q},\mathbf{s}}$ , obeying the following condition $|\Delta\omega_{\mathbf{q},\mathbf{s}}-i\Gamma_{\mathbf{q},\mathbf{s}}|<<\omega_{\mathbf{q},\mathbf{s}}$ [5,6,27], where $\Delta\omega_{\mathbf{q},\mathbf{s}}=\tilde{\omega}_{\mathbf{q},\mathbf{s}}-\omega_{\mathbf{q},\mathbf{s}}$, with $\omega_{\mathbf{q},\mathbf{s}}$ being the harmonic frequency. The power spectrum of a well defined phonon quasi-particle, $G_{\mathbf{q},\mathbf{s}}(\omega)$, should have a Lorentzian-type line shape with a phonon line-width equal to approximately $1/2\tau_{\mathbf{q},\mathbf{s}}$ [5, 6, 27], $\tau_{\mathbf{q},\mathbf{s}}$ being the $(\mathbf{q},\mathbf{s})$-mode lifetime.

We carried out Born-Oppenheimer MD (BOMD) simulations [28,29] on a 2x2x2 supercell of MgPv containing 160 atoms at the zero-pressure equilibrium volume for a series of temperatures from 100 K to 1,500 K. The harmonic normal modes were obtained using density functional perturbation theory (DFPT) [10] (see EPAPS for simulation details). The power spectrum, $G_{\mathbf{q}}(\omega)$, of MgPv at 700 K displays a single peak between 320 and 350 cm$^{-1}$ (Fig. 1a), which is in fact the superposition of three individual modes represented by $G_{\mathbf{q},\mathbf{s}}(\omega)$ 's (Fig. 1b). Therefore, analysis of $G_{\mathbf{q},\mathbf{s}}(\omega)$ allows unequivocal identification of $\tilde{\omega}_{\mathbf{q},\mathbf{s}}$ and $\Gamma_{\mathbf{q},\mathbf{s}}$ for each individual mode. Alternatively, one can obtain these quantities by fitting an exponentially decaying cosine function to the $(\mathbf{q},s)$-projected VAF (see EPAPS). For example, $\Delta\omega_{\mathbf{q},\mathbf{s}}$ for these modes are 6.1 cm$^{-1}$, -1.4 cm$^{-1}$, and 3.3 cm$^{-1}$, while their line-widths are 7.4 cm$^{-1}$, 8.3 cm$^{-1}$, and 7.7 cm$^{-1}$, respectively. These phonon quasi-particles are well defined according to the criterion indicated above [5, 6, 27].

MgPv is a weakly anharmonic system at mantle conditions [11-13], but anharmonic effects have been well demonstrated experimentally [14-16] at ambient pressure below 700 K for some Raman-active modes. At constant volume, the intrinsic temperature induced shifts of these modes, $\Delta\omega_{\mathbf{q},\mathbf{s}}$, exhibit diverse behaviors that can be classified as: $\Delta\omega_{\mathbf{q},\mathbf{s}}>0$, $\Delta\omega_{\mathbf{q},\mathbf{s}}<0$, and $\Delta\omega_{\mathbf{q},\mathbf{s}}\approx 0$ (Figs. 2a,b,c). The line-widths of these modes, $\Gamma_{\mathbf{q},\mathbf{s}}$, display strong temperature dependences, i.e., increase monotonically with temperature indicating that phonons, as expected, decay faster at higher temperature (Figs. 2g,h,i). Below 700 K, both $\Delta\omega_{\mathbf{q,s}}$ and $\Gamma_{\mathbf{q,s}}$ vary linearly with temperature, a behavior that can be captured by lowest order many-body perturbation theory (MBPT) [30,31]. Above 700 K such linear dependence no longer holds for those modes displaying the strong anharmonic behavior (Figs. 2a,g). Therefore, this strategy captures higher order anharmonic effects that cannot be quantified by MBPT.

Experimental validation of these results is important to establish this method as a legitimate approach to obtaining phonon quasi-particle properties and anharmonic effects on thermal properties. However, before comparing these results obtained at constant volume with data collected at constant pressure [14-16], we must correct for extrinsic frequency shifts caused by thermal expansion. At 0 GPa these shifts can be expressed as (see EPAPS):

$$\Delta\omega_{q,s}(T)\Big|_{P=0} = \Delta\omega_{q,s}(T)\Big|_{V=V_{eq}} + \omega_{q,s}\left(\exp(-\gamma_{q,s}\overline{\alpha}T)-1\right), \qquad (5)$$

where $\overline{\alpha}$ is the averaged thermal expansion coefficient, $\overline{\alpha}(T)=1/T\int_0^T \alpha(T')\,dT'$, $\alpha$ is the thermal expansion coefficient obtained with the QHA, and $\gamma_{\mathbf{q,s}}$ is the mode Grüneisen parameter obtained with DFPT [10]. Updated frequency shifts at 0 GPa are shown in Figs. 2d,e,f.

Table I shows a quantitative comparison of our results with experimental data [14] for $\gamma_{\mathbf{q,s}}$ and for the intrinsic shift, $\lambda_{\mathbf{q,s}}=\left(\dfrac{\partial\ln\widetilde{\omega}_{\mathbf{q,s}}}{\partial T}\right)_V$ (see EPAPS for an uncertainty analysis of experimental $\lambda_{\mathbf{q,s}}$). In overall, our results reproduce very well the temperature dependent behavior of the Raman frequencies below 700 K [14-16]: (i) with increasing temperature, $\Delta\omega_{\mathbf{q,s}}\big|_{P=0}$ increases for the 255 cm$^{-1}$ mode, stays nearly constant for the 245cm$^{-1}$ mode (Fig. 2d), and decreases for the other modes (Figs. 2e,f), as seen experimentally; (ii) the strongly anharmonic modes (245 cm$^{-1}$ and 255 cm$^{-1}$) have large positive $\lambda_{\mathbf{q,s}}$ while the other four weakly anharmonic modes have considerably smaller negative $\lambda_{\mathbf{q,s}}$'s; (iii) within uncertainties, there is very good agreement between predicted and measure intrinsic shifts ($\lambda_{\mathbf{q},\mathbf{s}}$). The agreement is more obvious for the strongly anharmonic modes, but differences between measured and calculated $\lambda_{\mathbf{q},\mathbf{s}}$'s are comparable for all these modes. These results also shed light on the origin of the frequency shifts. Usually, $\left.\dfrac{\partial \ln \tilde{\omega}_{q,s}}{\partial \mathrm{T}}\right|_P < 0$ because of the thermal expansion effect. Here we see that this justification applies only to the 384 and 508 cm$^{-1}$ modes (see Figs. 2c,f). Negative shifts displayed by the 283 and 395 cm$^{-1}$ modes are caused by combined effects of intrinsic anharmonicity and thermal expansion (see Figs. 2b,e).

The consistency between calculated and measured anharmonic effects validates this method and indicates that phonon-phonon interaction is sufficiently well accounted for in MD simulations using the 160-atom supercell. While this cell size seems sufficient to calculate renormalized phonon frequencies (see EPAPS), the small number of $\mathbf{q}$-vectors sampled in the BOMD supercell is not sufficient to converge calculations of thermodynamics properties, e.g., thermal expansivity or phase boundaries [3]. Free energy calculations with useful accuracy must not only use accurate $\tilde{\omega}_{\mathbf{q},\mathbf{s}}$'s but also sample them over a very large number of $\mathbf{q}$-points in the Brillouin zone corresponding to MD simulations using $\sim 10^4$ to $10^5$ atom supercells. Using BOMD, this is currently beyond the capability of direct free energy methods, such as thermodynamic integration (TI) [32]. However, large $\mathbf{q}$-sampling is possible if full anharmonic dispersions are available. The success of the QHA depends to a great extent on this thoroughly sampled Fourier interpolated phonon dispersions. Here we obtain full renormalized phonon dispersions from these calculated values of $\tilde{\omega}_{\mathbf{q},\mathbf{s}}$ using a similar interpolation scheme.

Introducing the renormalized "eigenvalue" matrix, $\Lambda_{\mathbf{q}}$, containing $\{\tilde{\omega}_{\mathbf{q},1}^2, \tilde{\omega}_{\mathbf{q},2}^2, ..., \tilde{\omega}_{\mathbf{q},3N}^2\}$ in the diagonal, and the matrix of harmonic eigenvectors, $[\hat{\mathbf{e}}_{\mathbf{q}}] = \{\hat{\mathbf{e}}_{1,\mathbf{q}}, ..., \hat{\mathbf{e}}_{3N,\mathbf{q}}\}$, we define an effective dynamical matrix,

$$\tilde{\mathrm{D}}(\mathbf{q}) = [\hat{\mathbf{e}}_{\mathbf{q}}]\Lambda_{\mathbf{q}}[\hat{\mathbf{e}}_{\mathbf{q}}]^+ . \qquad (6)$$

$\tilde{\mathrm{D}}(\mathbf{q})$ shares identical eigenvectors with the harmonic matrix, $\mathrm{D}(\mathbf{q})$, but has different eigenvalues. Therefore, anharmonic interactions are retained in the effective force constant matrix, $\widetilde{\Phi}(\mathbf{R})$, obtained by Fourier transforming $\widetilde{D}(\mathbf{q})$. As higher order interactions are usually shorter ranged compared to the harmonic ones, the calculation of $\widetilde{\Phi}(\mathbf{R})$ does not require BOMD supercells larger than those required for obtaining the harmonic $\Phi(\mathbf{R})$. Diagonalization of

$$\widetilde{D}(\mathbf{q}') = \sum_{\mathbf{R}} \widetilde{\Phi}(\mathbf{R}) \exp(-i\mathbf{q}'\mathbf{R}) \tag{7}$$

then gives $\widetilde{\omega}_{\mathbf{q}',s}$ for any $\mathbf{q}'$ in the Brillouin zone.

Renormalized phonon dispersions of MgPv obtained at $T = 1,500$ K are compared in Fig. 3a with phonon dispersions calculated using DFPT [10]. Frequencies obtained from a 160-atom BOMD simulation are compatible with those from DFPT for all high symmetry zone edge modes (see EPAPS and Fig. E2a for details). Anharmonic shifts are very discernible in the 1,500 K phonon dispersions and in the vibrational density of states (VDoS) (Fig. 3b) (for comparison see Fig. E2 in EPAPS). These anharmonic VDoS are invaluable for free energy calculations at very high temperatures when combined with PGM [4-7], where the entropy is analytically obtained from quasi-particle frequencies $\widetilde{\omega}_{\mathbf{q},s}$ [9]:

$$S = k_b \sum_{q,s} \left[ \frac{\hbar\widetilde{\omega}_{q,s}/k_bT}{\exp(\hbar\widetilde{\omega}_{q,s}/k_bT)-1} - \ln\left(1-\exp(-\hbar\widetilde{\omega}_{q,s}/k_bT)\right) \right]. \tag{8}$$

The free energy, $\mathbf{F}(V,T)$, is then obtained by numerically integrating $\mathbf{S}$ over T (see EPAPS). Fig. 4 shows a comparison between free energies computed using the PGM on different $\mathbf{q}$-grids and by TI [32] using results for the 2x2x2 BOMD supercell. Results obtained with the PGM and with the TI approach are equivalent for the same 2x2x2 $\mathbf{q}$-grid. However, there is a noticeable difference (~2% at 1500 K) for PGM results obtained on a densely interpolated $\mathbf{q}$-grid (10x10x8). Such difference is very significant for determining phase boundaries [33]. Formally, TI is an exact method for a given MD supercell. However, conducting converged TI using BOMD on a supercell sufficiently large (10x10x8 or 16,000 atoms for MgPv) is beyond current capability. Therefore, this strategy opens the door to an alternative route to achieving convergence in phase space sampling for accurate first-principles anharmonic free energy calculations.

In summary, we have overcome several obstacles in first principles computations of phonon quasi-particle properties and anharmonic free energy of crystalline systems. We reproduced irregular temperature induced frequency shifts [14-16] observed in the Raman spectrum of a weakly anharmonic and complex system, *Pbnm* $MgSiO_3$-perovskite. As long as phonon quasi-particles exist, i.e., as long as anharmonicity does not mix harmonic modes, and the power spectra of all modes have well defined single peaks, one can also obtain anharmonic dispersions over the entire Brillouin zone. Combination of these thoroughly sampled quasi-particle dispersions with the phonon gas model can offer first-principles free energy and thermodynamics properties in the thermodynamic limit (N $\rightarrow\infty$) at temperatures beyond the limit of validity of the QHA. This is especially important for calculations of thermal expansivity and phase boundaries [1,3,33], key properties in geodynamics simulations.

*Acknowledgments*: The authors thank Ann Hofmeister for helpful discussions. DBZ was supported primarily by the Abu Dhabi-Minnesota Institute for Research Excellence (ADMIRE). TS and RMW were supported by NSF grant EAR-1047629. Computational resources were provided by the Minnesota Supercomputing Institute and by XSEDE.

Table I. Comparison between measured [14] and calculated phonon frequencies, $\tilde{\omega}_{\mathbf{q,s}}$, mode Grüneisen parameter, $\gamma_{\mathbf{q,s}}$, and intrinsic mode anharmonicity parameters, $\lambda_{\mathbf{q,s}}$, for Raman active modes of MgPv at ambient conditions. Harmonic frequencies, $\omega_{\mathbf{q,s}}$, in parenthesis and $\gamma_{\mathbf{q,s}}$ are DFTP results [10] in italic. The detailed evaluation of experimental uncertainties in $\lambda_{\mathbf{q,s}}$ can be found in the EPAPS material.

| | $\tilde{\omega}_{\mathbf{q,s}}$ ($\omega_{\mathbf{q,s}}$) (cm$^{-1}$) | | $\gamma_{\mathbf{q,s}}$ | | $\lambda_{\mathbf{q,s}}$ (10$^{-5}$ K$^{-1}$) | |
|---|---|---|---|---|---|---|
| Sym. | This work | Exp.[14] | This work | Exp.[14] | This work | Exp.[14] |
| A$_g$ | 250 (*245*) | 252 | *2.01* | 2.21 | 5.91±0.23 | 6.25±1.66 |
| B$_{2g}$ | 263 (*255*) | 257 | *2.40* | 2.12 | 10.47±0.06 | 9.20±3.69 |
| A$_g$ | 281 (*283*) | 281 | *1.38* | 1.26 | -2.58±0.22 | -0.65±0.98 |
| A$_g$ | 393 (*395*) | 390 | *1.38* | 1.39 | -1.73±0.04 | -0.85±1.25 |
| A$_g$ | 384 (*384*) | 379 | *0.97* | 1.85 | -0.37±0.02 | -0.28±0.96 |
| B$_{3g}$ | 507 (*508*) | 501 | *1.62* | 1.71 | -0.45±0.03 | -0.42±0.69 |

**Figure Captions:**

**Figure 1**. (color online) (a) Power spectrum, $G_{\mathbf{q}}$, of Mg-Pv (red solid line) calculated from a BOMD run at 700 K in a supercell containing 160 atoms. ($\mathbf{q}$=0) in $\sum_{\mathbf{s}} G_{\mathbf{q},\mathbf{s}}$ (blue dashed line) correspond to modes with frequencies (harmonic frequencies) 329 cm$^{-1}$ (323 cm$^{-1}$), 333 cm$^{-1}$ (334 cm$^{-1}$), and 343 cm$^{-1}$ (340 cm$^{-1}$). (b) Mode-projected power spectra, $G_{\mathbf{q},\mathbf{s}}$, of these modes. $\sum_{\mathbf{s}} G_{\mathbf{q},\mathbf{s}}$ resembles $G_{\mathbf{q}}$ in the shaded area.

**Figure 2**. (color online) (a), (b), and (c) show the temperature dependent frequency shifts at constant volume, $\Delta\omega_{q,s}\mid_{V(P=0)}$, of MgPv modes with positive, negative, and nearly zero shifts, respectively. (d), (e), and (f) show frequency shifts at zero pressure $\Delta\omega_{q,s}\mid_{P=0}$ given by Eq. (6) (see EPAPS for details). The temperature dependence of these Raman-active modes has been experimentally measured [14-16] at ambient pressure. See Table I for comparison. (g), (h), and (i) show temperature dependent line widths, $\Gamma_{\mathbf{q},\mathbf{s}}$, at constant volume for the same modes, respectively. The volume is equal to that of a static lattice with (LDA) pressure equal to 0 GPa. Solid lines indicate linear behavior, $\Delta\omega_{\mathbf{q},\mathbf{s}} \approx \alpha_{\mathbf{q},\mathbf{s}} T$ and $\Gamma_{\mathbf{q},\mathbf{s}} \approx \beta_{\mathbf{q},\mathbf{s}} T$, for $T < 700$ K. Error bars represent uncertainties originating in BOMD runs.

**Figure 3**. (color online) (a) Anharmonic phonon dispersions (blue solid lines) of Mg-Pv obtained at $T = 1,500$ K. Harmonic phonon dispersions calculated using DFTP at the same volume (red dashed lines) are shown for comparison. (b) Temperature dependent VDoS of Mg-Pv obtained on a 10x10x8 $\mathbf{q}$-point grid using DFPT (red dashed line) and at 1,500K (blue solid line), both at the same volume. The thermal pressure is ~10 GPa. (See EPAPS for more detailed comparisons and simulation details.)

**Figure 4**. (color online) Vibrational free energy, $F$, and (insert) excess anharmonic vibrational free energy ($F_A$) and entropy ($S_A$) obtained with the PGM and the TI approach using an MD run sampling a coarse $\mathbf{q}$-grid (2x2x2) (see EPAPS). Results with a finely interpolated $\mathbf{q}$-grid (10x10x8) use the PGM. Error bars are evaluated from 5 parallel replicas (60 ps each) of BOMD.

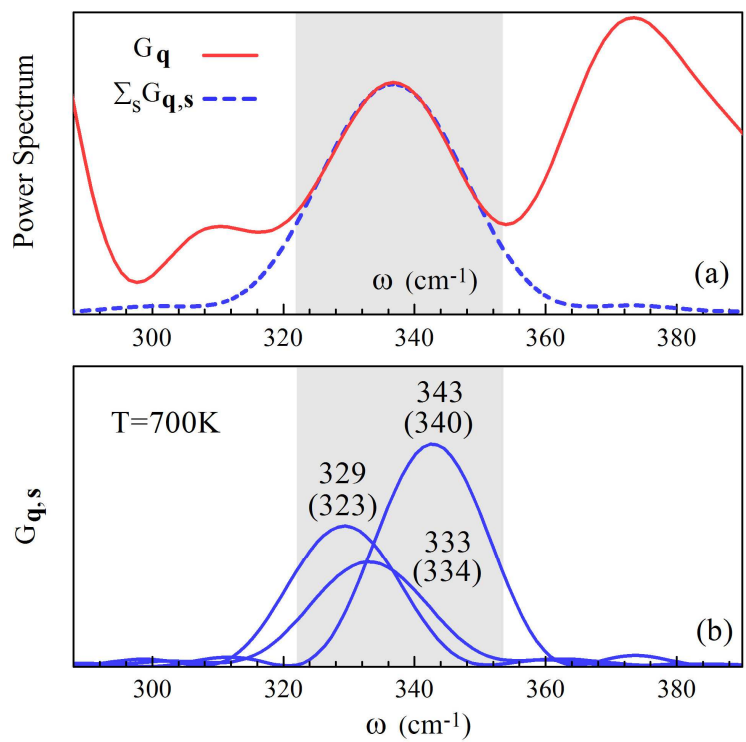

**Figure 1**

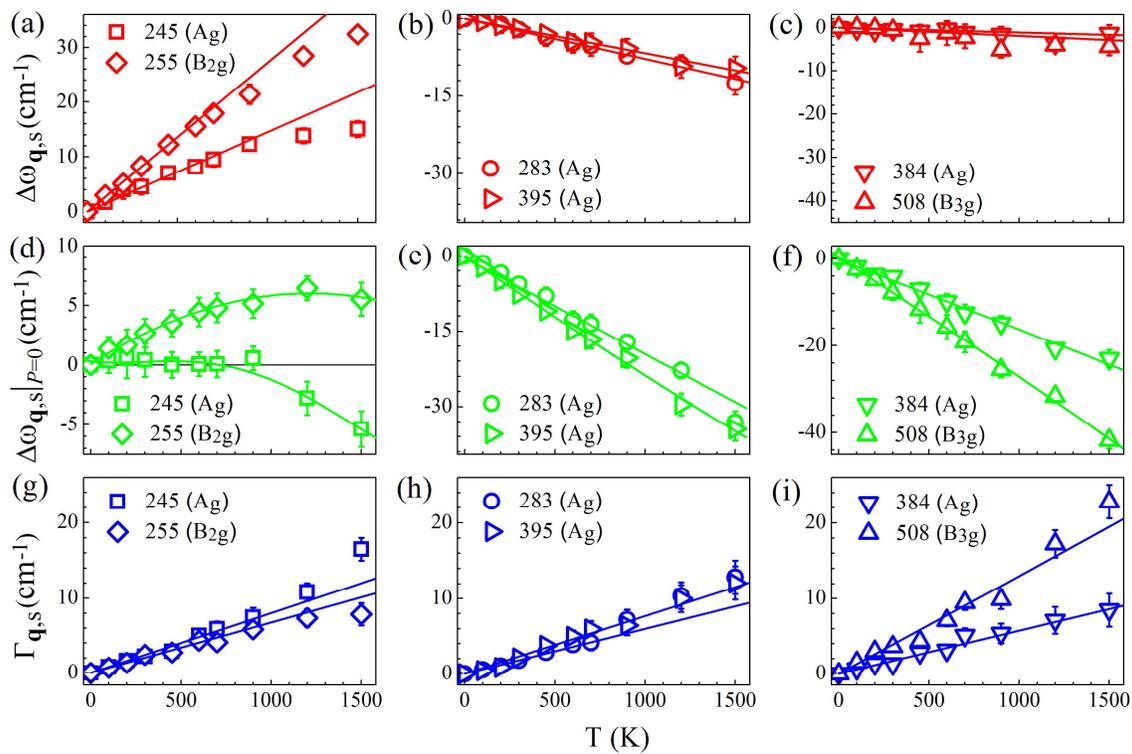

**Figure 2**

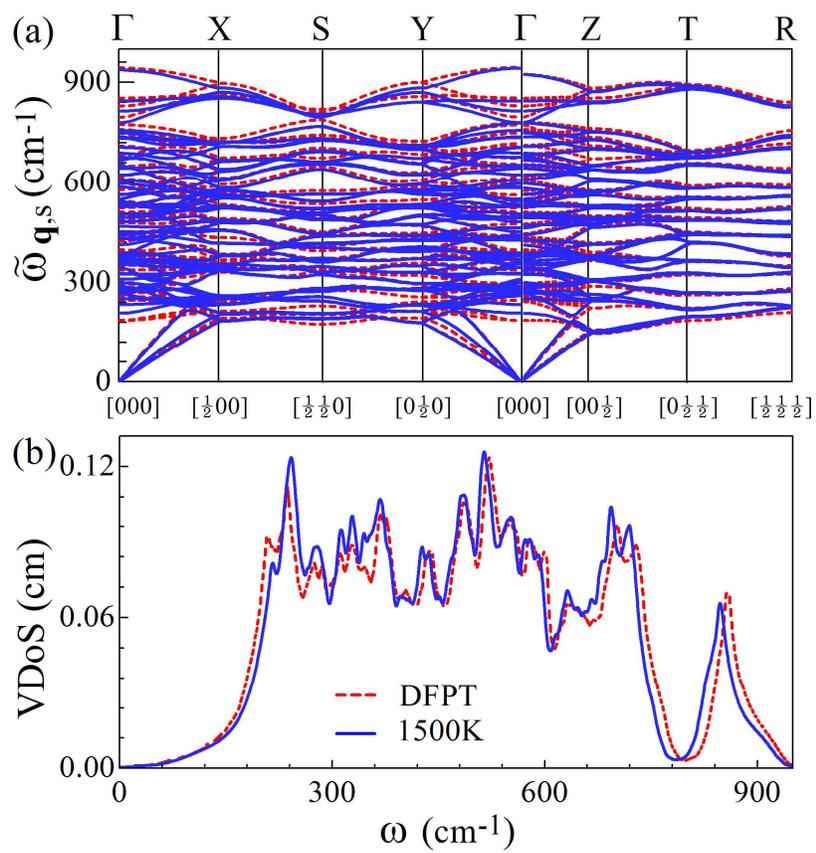

Figure 3

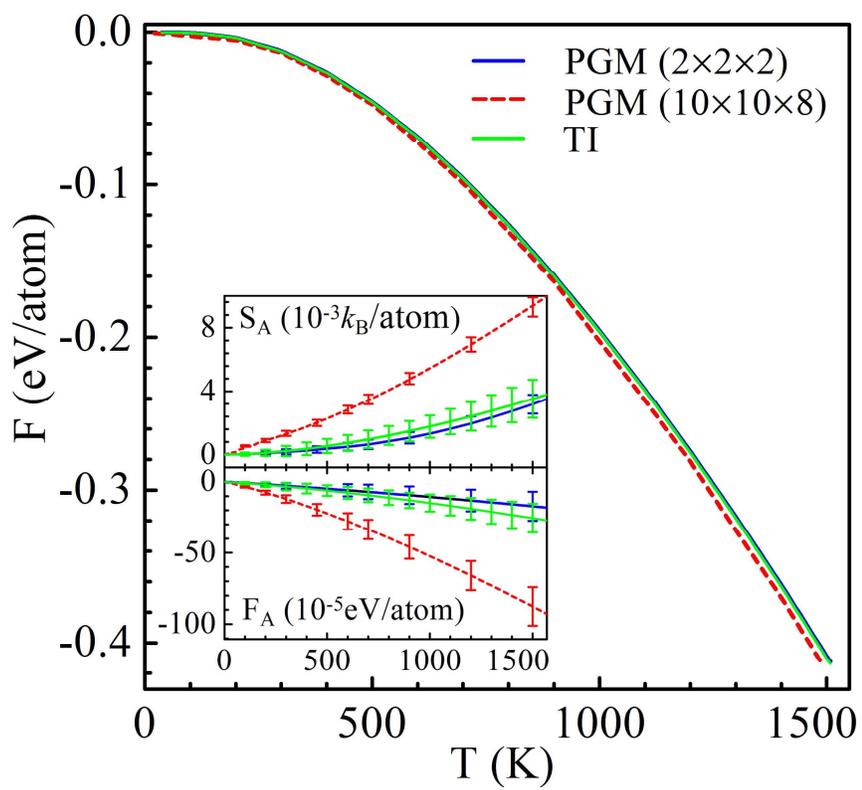

**Figure 4**